\documentclass[fleqn,10pt]{wlscirep}
\usepackage[utf8]{inputenc}
\usepackage[T1]{fontenc}
\usepackage{float}
\usepackage{subcaption}
\usepackage{breqn}
\title{Unconventional quantum criticality in a non-Hermitian extended Kitaev chain}

\author[1,2,3]{S Rahul}
\author[4,5]{Nilanjan Roy}
\author[1,2]{Ranjith R Kumar}
\author[1,2]{Y R Kartik}
\author[1,*]{Sujit Sarkar}
\affil[1]{Department of Physics, Indian Institute of Technology Madras, Chennai 600036, India}
\affil[2]{Theoretical Sciences Division, Poornaprajna Institute of Scientific Research, Bidalur, Bengaluru 562164, India.}
\affil[3]{Graduate Studies, Manipal Academy of 
	Higher Education, Madhava Nagar, Manipal-576104, India.}
\affil[4]{Department of Physics, Indian Institute of Science Education and Research, Bhopal, Madhya Pradesh 462066, India}	
\affil[5]{Centre for Condensed Matter Theory, Department of Physics, Indian Institute of Science, Bangalore 560012, India}

\affil[*]{corresponding author: sujit.tifr@gmail.com}

\begin{abstract}
We investigate the nature of quantum criticality and  topological phase transitions near the critical lines obtained for the extended Kitaev chain with next nearest neighbor hopping parameters and non-Hermitian chemical potential. We surprisingly find multiple gap-less points, the locations of which in the momentum space can change along the critical line unlike the Hermitian counterpart. The interesting simultaneous occurrences of vanishing and sign flipping behavior by real and imaginary components, respectively of the lowest excitation is observed near the topological phase transition.  
Introduction of non-Hermitian factor leads to an isolated critical point instead of a critical line and hence, reduced number of multi-critical points as compared to the Hermitian case. The critical exponents obtained for the multi-critical and critical points show a very distinct behavior from the Hermitian case.
\end{abstract}
\begin{document}

\flushbottom
\maketitle
%
%
\thispagestyle{empty}

\section{Introduction}

The study of topological properties of non-Hermitian systems has acquired a growing attention in the recent times since these systems possess a rather unique set of features such as complex excitation spectrum \cite{shen2018topological}, special degenerate points called the exceptional points (EPs) \cite{bergholtz2019exceptional}, modification of bulk boundary correspondence \cite{koch2020bulk,PhysRevLett.121.026808} etc. 
Topological phases in the non-Hermitian systems have been shown to host Majorana zero modes (MZMs)\cite{yuce2016majorana}.
Recent studies have shown that non-Hermitian Hamiltonains support topological non-trivial phases, for example, one dimensional Kitaev chain with the complex chemical potential shows topologically trivial and non-trivial phase with MZMs  \cite{zeng2016non}. 
Also fractional invariant number has been realized in the non-Hermitian system with anisotropy in the hopping \cite{yin2018geometrical}. 
The non-Hermitian Hamiltonians which obey $PT$ symmetry with real eigenvalues \cite{bender1998real,bender2007making} undergo transition from $PT$ symmetric to $PT$ symmetry broken region through exceptional points which are degenerate points \cite{bender2007making}. 
$PT$ symmetric Hamiltonians have been explored in one dimensional topological systems \cite{yao2018edge,yokomizo2019non,lieu2018topological,zhu2014pt,leykam2017edge} and it has got a variety of application in quantum information \cite{croke2015pt}.     
Topological non-Hermitian systems have found applications in various areas such as, non-equilibrium open quantum systems \cite{el2018non,kozii2017non}, correlated electronic systems  \cite{yoshida2018non,bergholtz2019non} and
in the novel lasing techniques enabled by edge states amplification which is topologically protected against disorder and defects \cite{st2017lasing,parto2018edge,harari2018topological}. 
Non-Hermitian systems can be thought of as a Hermitian systems interacting with the environment which is represented by the non-Hermitian term in the Hamiltonian.
There have been numerous studies where one dimensional Hermitian topological systems have been connected to an external bath \cite{klett2017relation} to understand the stability of MZMs, symmetry and topological phase transition.        
 \begin{figure}
	\centering
	\includegraphics[width=9cm,height=9cm]{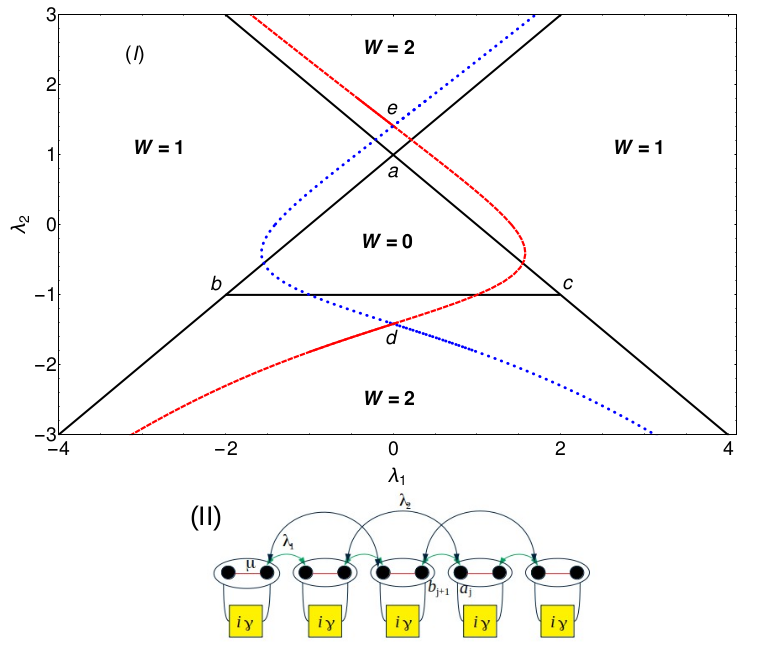}
	\caption{(I) Phase diagram of the non-Hermitian model Hamiltonian in presence of $\gamma=1$ and $\mu=1$. Dashed and dotted lines corresponds to non-Hermitian critical lines. Solid black lines corresponds to Hermitian critical lines where $\gamma=0$. (II) A schematic representation of the model Hamiltonian in presence of non-Hermitian factor $\gamma$ introduced in the chemical potential. $a_j$ and $b_{j}$ are the Majorana operators where fermionic operator $c_j=(a_j + ib_j)/2$. $\lambda_1$, $\lambda_2$ and $\mu$ are the nearest neighbor, next nearest neighbor hopping parameters and on-site chemical potential respectively.}
	\label{sch}
\end{figure}
These non-Hermitian systems have been
studied in the context of dissipative quantum systems \cite{PhysRevLett.102.065703,diehl2011topology}, and most recently in topological insulators
\cite{lee2016anomalous,leykam2017edge} and topological superconductors \cite{wang2015spontaneous,san2016majorana}.\\
In this work, we study the role of non-Hermiticity on the topological phases and near phase transitions in an extended Kitaev chain. 
To be more specific, we investigate the complex excitation spectrum and obtain the critical lines of the non-Hermitian extended Kitaev chain with next nearest neighbor hopping parameters. We study the gap closing points in the excitation spectra from the analysis of k. We also calculate the critical exponents for the multicritical points and critical lines of the non-Hermitian model and discuss the difference in behavior between Hermitian and non-Hermitian cases. \vspace{0.25cm} \\
\textbf{Model:}\\
We consider an extended Kitaev chain in presence of non-Hermiticity given by,
\begin{eqnarray}
	H =&& -(\mu+i \gamma) \sum_{i=1}^{N} (1 - 2 c_{i}^{\dagger}c_{i}) - \bigg[\lambda_1 \sum_{i=1}^{N-1} (c_{i}^{\dagger}c_{i+1} + c_{i}^{\dagger}c_{i+1}^{\dagger}) + \lambda_2 \sum_{i=2}^{N-1} ( c_{i-1}^{\dagger}c_{i+1} +  c_{i+1} c_{i-1}) + h.c.\bigg].
	\label{jw1} 
\end{eqnarray} 
Where $\lambda_1$, $\lambda_2$, $\mu$ and $\gamma$ corresponds to nearest, next nearest neighbor hopping parameters, chemical potential and non-Hermitian factor respectively. The inclusion of the imaginary term $i\gamma$ to the extended Kitaev model is graphically depicted in the Fig.\ref{sch}(II). The non-Hermitian factor $\gamma$ acts as a local bath at each site of the chain. Here $c^\dagger_i$ is the spinless fermionic creation operator at site $i$.
After the Fourier transformation, the Hamiltonian in Eq.~\ref{jw1} becomes, 
\begin{eqnarray}
	H =&& \sum_{k} (2(\mu+i \gamma) - 2 \lambda_1 \cos k - 2 \lambda_2 \cos 2k) c_{k}^{\dagger} c_{k} + i \sum_{k} (2 \lambda_1 \sin k c_{k}^{\dagger}c_{-k}^{\dagger} + 2 \lambda_2 \sin 2k c_{k}^{\dagger}c_{-k}^{\dagger} + h.c). 
\end{eqnarray}
The Bogoliubov-de-Gennes (BdG) form of the Hamiltonian is,
\begin{eqnarray}
	H_k =  \chi_{z} (k) \sigma_z - \chi_{y} (k) \sigma_y = \left(\begin{matrix}
		\chi_{z} (k) && i\chi_{y} (k)\\
		-i\chi_{y} (k) && -\chi_{z} (k)\\
	\end{matrix} \right),
	\label{APS}
\end{eqnarray}
where $ \chi_{z} (k) = -2 \lambda_1 \cos k - 2 \lambda_2 \cos 2k + 2(\mu+ i \gamma),$ and $ \chi_{y} (k) = 2 \lambda_1 \sin k + 2 \lambda_2 \sin 2k$. 
The energy dispersion relation is given by,
\begin{eqnarray}
	E=\pm \sqrt{(\chi_{z} (k))^2 + (\chi_{y} (k))^2}.
	\label{es} 
\end{eqnarray}    
Since the energy eigenvalues are complex throughout the parameter space, model Hamiltonian Eq.~\ref{APS}  does not obey $PT$ symmetry which can be verified using the $PT$ symmetry operator $\hat{K} \sigma_z$ acting on the Hamiltonian $H$, i.e., $\hat{K} \sigma_z \hat{H_k}  \sigma_z \hat{K^{-1}} \ne \hat{H_k} $. 
The model Hamiltonian possess three topological phases characterized by winding number $W=0$, $1$ and $2$ which are separated by the critical lines presented in the Fig.\ref{sch}(I). The solid black lines in the Fig.\ref{sch}(I) are the Hermitian critical lines which is presented to differentiate the Hermitian and non-Hermitian criticality. In this case the topological cases are characterized using two ways. One is by calculating the winding number using the expression, 
\begin{eqnarray}
	W = \frac{1}{2\pi} \oint F dk, 
\end{eqnarray}
where F is the curvature function defined in the Eq.\ref{cf1}.
It is possible to define the integer valued winding number to a non-Hermitian system in the presence of a complex chemical potential. There are also evidences to define topological phases through complex Berry phase in 1D \cite{2304.12723}. In the presence of complex potential, the Hamiltonian can be expressed in biorthnormal vectors, where the exceptional points are represented by eq 6. The effective winding number can be  calculated by integrating the curvature function around the exceptional points \cite{yin2018geometrical,aquino2023critical}. Transition can also be recognized by the complex energy spectrum, where the energy gap closes and reopens during this process signalling a topological phase transition. This kind of non-Hermitian system posses a line gap in the complex spectrum, where the winding numbers can be mapped to their Hermitian counterparts. And moreover, there are no evidences of non-Hermitian skin effects in this kind of models \cite{PhysRevLett.124.086801}. 
The other method is by calculating the zeros of the Hamiltonian by writing it in as complex function \cite{verresen2018topology,rahul2022topological}. Here winding number expression can be written in terms of zeros and poles of a complex function which takes the form, $W= z-p$, where z is the number of zeros and p is number of poles. In our case since there are poles, the winding number is directly equals to number of zeros of the complex function.
We attempt to characterize all the topological phases and criticality using the zero mode solutions (ZMS) method. For the Hermitian case, the critical lines are fairly easy to obtain since there is no complex terms to deal with but at criticality we use the same method "ZMS" to characterize and obtain the topological invariant number \cite{rahul2021majorana}.\\

\begin{figure}[H]
	\centering
	\includegraphics[width=9cm,height=9cm]{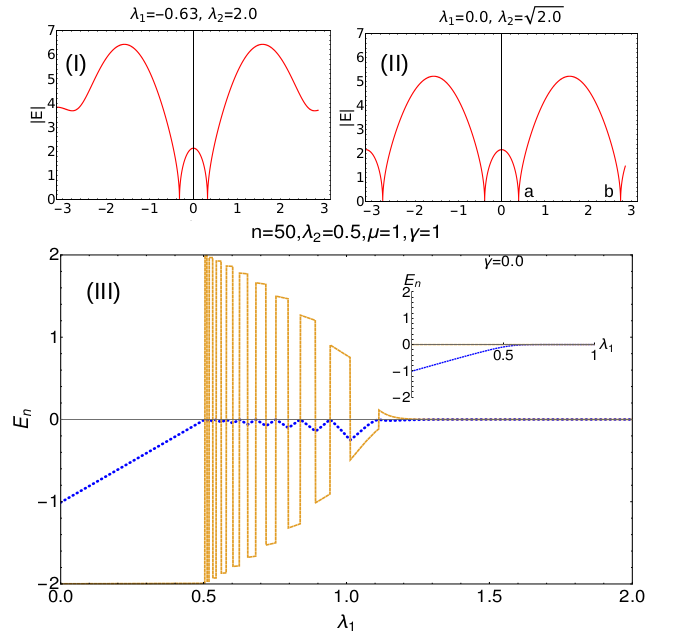}
	\caption{(I) and (II) Energy spectrum $|E|$ is plotted with respect to $k$ at a critical and multicritical point (e) respectively. a and b marked in the plot (II) are the two gap closing points at the multicritical point e. (III) Eigenvalues plotted with respect to the parameter $\lambda_1$ in the open boundary setting with system size $n=50$. Dotted and dashed curves represents real and imaginary part of the zero mode eigenvalue respectively. Inset corresponds to the Hermitian case ($\gamma = 0$).}
	\label{peri}
\end{figure}

\noindent{\textbf{Results}}\\
{\textbf{Multiple gap closing phenomena:} \label{period}}
Here we analyze the excitation spectra for topological phase transitions between the topological phases, $W=0$ to $1$ and $W=1$ to $2$. Comparing the Hermitian and non-Hermitian versions of the model Hamiltonian, we observe a striking difference where the third critical line $(\lambda_2 = \mu - \lambda_1)$ degenerate isolated critical point in the non-Hermitian case where the critical line $(\lambda_2 = \mu - \lambda_1)$ in the Hermitian case is a phase boundary between the topological phases $W=0$ and $W=2$.  
In the non-Hermitian case, the third critical line becomes an isolated critical point becoming degenerate with the critical line $\lambda_2 = \mu - \lambda_1$ at position $\lambda_1 = 0$, $\lambda_2 = \sqrt{\mu^2 + \gamma^2}$. Hence we study the $W=0$ to $1$, $W=1$ to $2$ topological phase transitions and later we study the behavior of the multicritical and critical points.\\
In the Fig.~\ref{peri}(I) and Fig.~\ref{peri}(II), we study the behavior of gap closing points in the excitation spectrum along the critical lines.\\

Along the dashed critical line (Fig.\ref{sch}), gap closing points in the excitation spectra continuously changes. The change in the gap closing points is governed by the Eq.\ref{kk}.
Real and imaginary part of the zero mode eigenvalue is studied with respect to the system parameter $\lambda_1$. In the Hermitian case the gap closing points corresponding to three critical lines are at $k=0, \pm \pi$ and $ \cos^{-1}(-\lambda_1/2\lambda_2)$ respectively~\cite{niu2012majorana}. These gap closing points do not change along the respective critical lines.

In the non-Hermitian case the values of $k$ at which the gap closes keep changing indicating the conceptual difference in the criticality between the non-Hermitian model and its Hermitian counterpart. In Fig.\ref{peri}(III) we show the behavior of real and imaginary eigenvalues in the open boundary setting. The real part (dotted) shows fluctuatory behavior whereas the imaginary part (solid) goes from negative to positive energy axes or vice versa when the real part goes to zero. This behavior starts at the Hermitian critical point and persists till the non-Hermitian critical point where both real and imaginary parts go to zero. In other way, this interesting behavior of zero mode eigenvalue is observed in the region between the Hermitian and non-Hermitian critical points. We study only the zero mode eigenvalue to avoid the complexity in analyzing the bulk eigenvalues. The behavior of the real and imaginary parts of the energy eigenvalue has been analyzed analytically for periodic boundary condition in the Methods section. The real and imaginary eigenvalues behave in the same way for all other topological phase transitions.

\begin{figure}[H]
	\centering
	\includegraphics[scale=0.18]{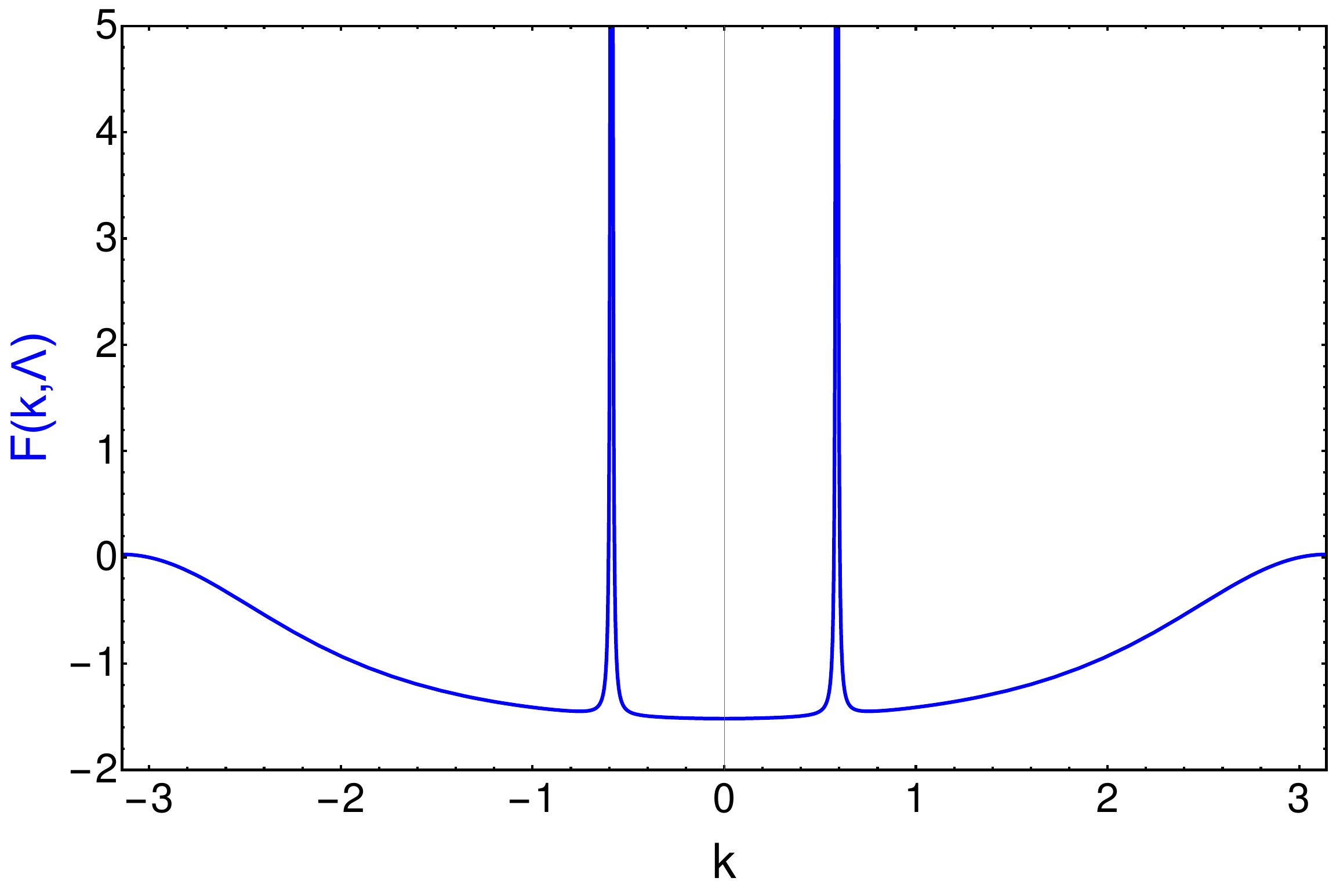}
	\caption{Curvature function $F(k,\Lambda)$ plotted with respect to $k$ at a critical point for the fixed values of $\lambda_1=0.970$, $\lambda_2=0.5$, $\mu = 1$ and $\gamma = 1$.} 
	\label{dispCF}
\end{figure}

\begin{figure}
	\centering
	\includegraphics[width=12cm,height=9cm]{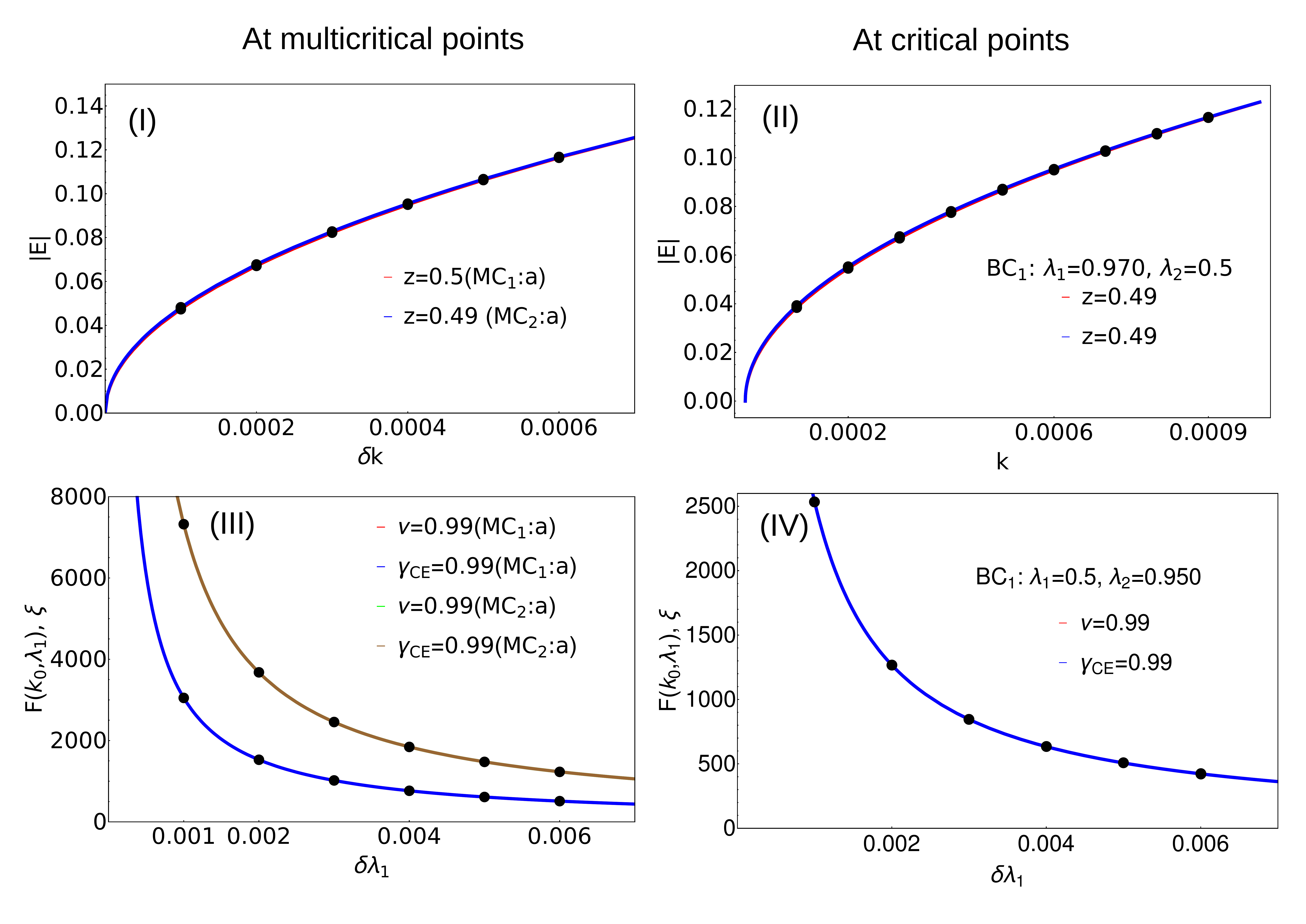}
	\caption{Critical exponents $z$, $\nu$ and $\gamma_{CE}$ calculated at the multicritical points $MC_1$ (e) and $MC_2$ (d). (I) and (III): Critical exponents $z$, $\nu$ and $\gamma_{CE}$ are calculated at the gap closing point $a$ of multicritical points $MC_1$ (e) and $MC_2$ (d). (II) and (IV): Critical exponents $z$, $\nu$ and $\gamma_{CE}$ are calculated at the critical point $BC_1$.}
	\label{mc1}
\end{figure}

\begin{figure}[H]
	\centering
	\includegraphics[width=12cm,height=9cm]{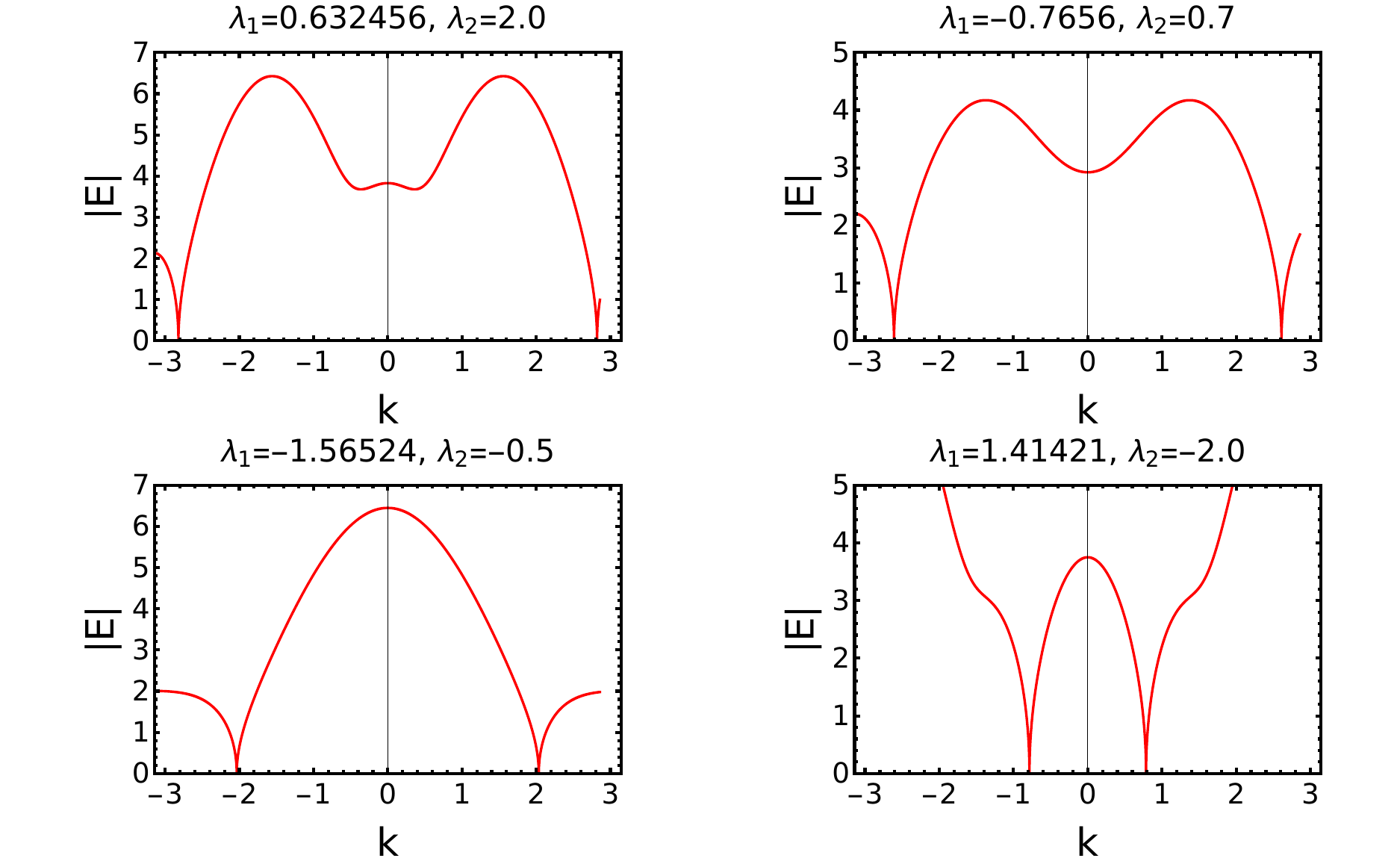}
	\caption{Behavior of absolute energy spectra (under periodic boundary conditions) along the dotted critical line in phase diagram Fig.\ref{sch}(I).}
	\label{periodb}
\end{figure}

\begin{figure}[H]
	\centering
	\includegraphics[width=6cm,height=4cm]{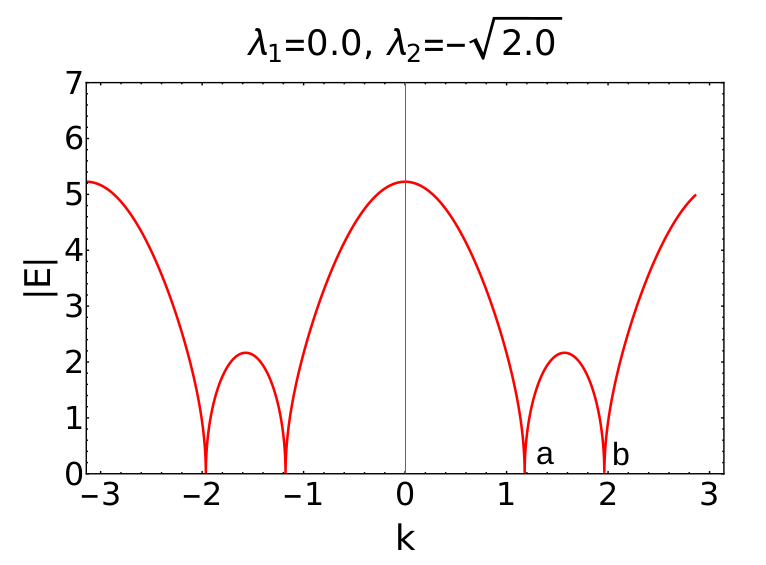}
	\caption{Excitation dispersion $|E|$ plotted with respect to $k$ at a multicritical point "d"(Fig.\ref{sch}(I)).}
	\label{MC2}
\end{figure}

\begin{figure}[H]
	\centering
	\includegraphics[width=6cm,height=4cm]{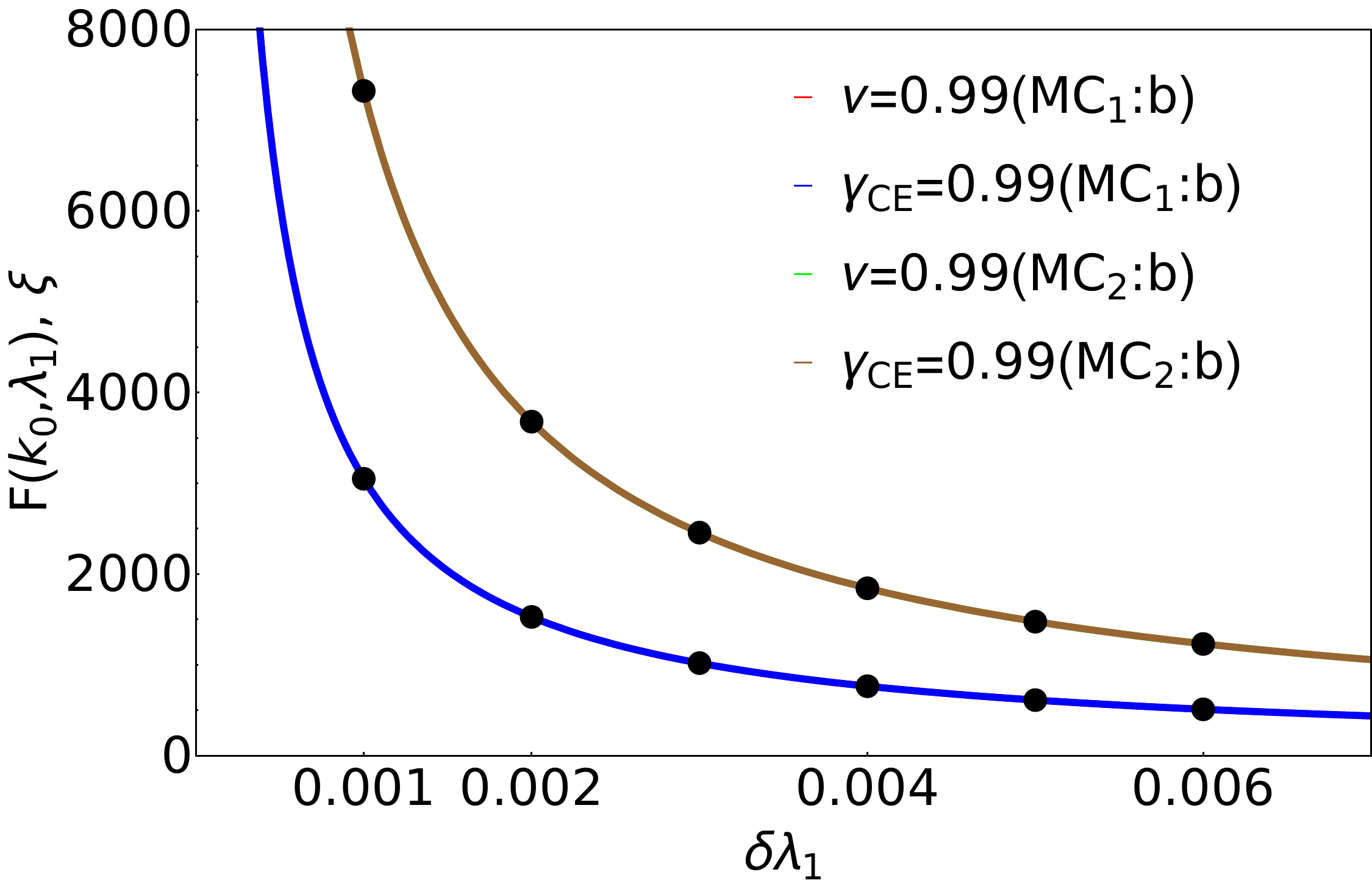}
	\caption{Critical exponents $\nu$ and $\gamma_{CE}$ calculated at the multicritical points $MC_1$ (e) and $MC_2$ (d).}
	\label{crit}
\end{figure}

By equating the Eq.~\ref{es} to zero we obtain an analytical expression for $k_{\pm}$ which gives the values of $k$ at the gap closing points respectively. The analytical expression of $k_{\pm}$ is given by,
\begin{dmath}
	k_{\pm} = \eta_1 \arccos \left[\frac{1}{4 \lambda_2 (\gamma - i \mu)} \left(-\gamma \lambda_1 - i \lambda_1 \lambda_2 + i \lambda_1 \mu + \eta_2 \sqrt{-(i \gamma + \lambda_2 + \mu)^2 (\lambda_1^2 + 4 i \gamma \lambda_2 + 4 \lambda_2 \mu)} \right)\right],
	\label{kk}
\end{dmath}
where $\eta_1$ and $\eta_2$ take different signs, i.e., $(-,-)$, $(-,+)$, $(+,-)$ and $(+,+)$ respectively which makes four solutions of $k$. 
At any critical point, $k_{(-,+)}$ and $k_{(+,+)}$ equations provide the information on the the gap closing location in the excitation spectra (Fig.~\ref{peri}(I)). Similarly at the multicritical points, all four equations of k provides the location of the gap closing in the excitation spectra (Fig.~\ref{peri}(II)). 
The excitation spectrum and $k$ equations along the dotted critical line (Fig.\ref{sch}) also shows similar behavior which is presented in the Fig.~\ref{periodb} in Methods section. The gap closing points along the dotted critical line is also governed by the $k$ expression Eq.\ref{kk}.\\ 
For studying the critical exponents such as $\gamma_{CE}$ and $\nu$ the divergent property of the curvature function at criticality is essential. The behavior of the curvature function is presented in the Fig.\ref{dispCF}. At the gap closing points in the excitation spectra, the divergence of the curvature function $F$ is observed. In the non-Hermitian case too, the curvature function $F$ retains the divergent behavior at the criticality and thus the critical exponents can be calculated which is studied in the next section. 
\noindent{\textbf{Critical exponents for multicritical points:}}
The dynamical critical exponent $z$, which takes the form $|E| \propto k^z$. Depending on the value of the dynamical critical exponent, the nature of the excitation dispersion. The critical exponents $\nu$ and $\gamma_{CE}$ are defined with the consideration of the curvature function. These critical exponents are calculated numerically using curve fitting method where the fitting curve takes the Orenstien-Zernike \cite{PhysRevB.95.075116} form given as, 
\begin{equation}
	F(k_0+\delta k, \Lambda_c) = \frac{F(k_0,\Lambda)}{1+\xi^2 \delta k^2 + \xi_c^4 \delta k^4},
\end{equation}
with the divergent behavior: $F(k_0,\Lambda) \propto |\Lambda - \Lambda_c|^{-\gamma_{CE}}$ and $\xi \propto |\Lambda - \Lambda_c|^{-\nu}$.\\  
In the topological systems, multicritical point separates more than two topological phases. Considering the example of Hermitian extended Kitaev chain, it has three topological phases, $W=0$, $1$ and $2$ and two multicritical points at $\lambda_1 = 0, \lambda_2 = 1$ and $\lambda_1 = 2, \lambda_2 = -1$. Authors of the reference \cite{kumar2021multi} have studied the properties of both the multicritical points. 
With the same spirit of investigation we study the nature of dispersion and its critical exponents for the two multicritical points present in the non-Hermitian systems. 
We have observed an interesting behavior in the nature of multicritical points in the non-Hermitian model compared to the Hermitian case. These multicritical points in the non-Hermitian case are displaced to its new locations, $\lambda_1 = 0$, $\lambda_2 = \sqrt{\mu^2 + \gamma^2}$ (e) and $\lambda_1 = 0$, $\lambda_2 = -\sqrt{\mu^2 + \gamma^2}$ (d) (For the points "e" and "d", refer the phase diagram presented in the Fig.\ref{sch}(I)) in contrast with the Hermitian case. The gap closing points in the excitation spectra for multicritical point $MC_1$ (e) is presented in the Fig.\ref{peri}(II). The excitation spectra for the multicritical point $MC_2$ (d) is shown in Fig.\ref{MC2} which is presented in the Methods section.\\
We calculate the critical exponents for the multicritical and critical points which are shown in the Fig.\ref{mc1}. 
The non-Hermitian case hosts two multicritical points $\lambda_1 = 0$, $\lambda_2 = \sqrt{\mu^2 + \gamma^2}$ (e) and $\lambda_1 = 0$, $\lambda_2 = -\sqrt{\mu^2 + \gamma^2}$ (d) as shown in the Fig.\ref{sch}(I). Both these multicritical points possess same critical exponents and nature of dispersion.  
In the Hermitian case, one of the multicritical point has linear dispersion dynamical critical exponent $z=1$ and the other multicritical point has quadratic dispersion with $z=2$ \cite{kumar2021multi,PhysRevB.107.205114}.
In the non-Hermitian model, the dynamical critical exponent $z$ acquires the values $0.49$ and $0.5$ at the gap closing point a of $MC_1$ (e) and $MC_2$ (d). By investigating the nature of excitation spectra through out the phase diagram, the dynamical critical exponent $z$ acquires the value $0.5$ which is discussed in the plots~(I) and (II) for both multicritical point and critical line. The dynamical critical exponent $z$ acquiring the value $0.5$ throughout the criticality shows a major difference between the Hermitian and the non-Hermitian cases. For $\gamma = 0$, i.e., in the Hermitian case, the authors of \cite{kumar2021multi} have shown that the Lorentz invariance is preserved in one of the multicriticality and is violated in the other due to the quadratic nature of the dispersion \cite{PhysRevB.100.195432}. In the non-Hermitian case, at both the multicriticality, the dynamical critical exponent $z$ takes the value $0.5$. 
The critical exponents $\nu$ and $\gamma_{CE}$ are calculated using the curvature function $F(k,\Lambda)$ where $\Lambda$ takes the system parameters $\lambda_1$ and $\lambda_2$ that are varied. 
The curvature function of the model Hamiltonian is given by, 
\begin{equation}
	F = (\partial_k \phi_1 + \partial_k \phi_2),
	\label{cf1} 
\end{equation}
where, \begin{equation}
	\phi_1 = \arctan\left( \frac{2 \lambda_1 \sin k + 2 \lambda_2 \sin 2k + 2\gamma}{-2 \lambda_1 \cos k - 2 \lambda_2 \cos 2k + 2\mu} \right), 
\end{equation} 
and 
\begin{equation}
	\phi_2 = \arctan\left( \frac{2 \lambda_1 \sin k + 2 \lambda_2 \sin 2k - 2\gamma}{-2 \lambda_1 \cos k - 2 \lambda_2 \cos 2k + 2\mu} \right). 
\end{equation}
The general notion is that, as one approaches the critical point, the curvature function diverges at the respective k value. In Hermitian systems, the critical lines are calculated for $k=0,\pm \pi$ and can take more values of k depending on the number of hopping parameters present in the system. An important contrast in the non-Hermitian model under consideration is that the gap closing no longer occurs at $k=0,\pm\pi$ which is also one of the primary reasons for the difficulty in calculating critical lines.  
But the presence of multiple gap closing points is associated with the divergence of the curvature function and 
hence the critical exponents can be calculated without any hindrance.\\ 
The calculation of critical exponents $\nu$ and $\gamma_{CE}$ at the multicritical point and critical point is discussed in Fig.\ref{mc1}(III) and Fig.\ref{mc1}(IV). The calculation of critical exponents at other gap closing points are discussed in detail in the Fig.\ref{crit} presented in the Methods section.\\

\noindent{\textbf{Critical exponents for critical points:}}\\
The behavior of the critical points is as interesting as that of the multicritical points which is presented in the Fig.\ref{mc1}(II) and Fig.\ref{mc1}(IV). 
According to the critical exponents calculations at several critical points on the criticality, the dynamical critical exponent acquires the value $0.5$ throughout the phase diagram which marks a significant difference from the Hermitian case \cite{kumar2021multi}. In the non-Hermitian case, the nature of the spectra remains the same at all the criticalities and also validate the scaling law which is explicitly explored in \cite{kumar2021multi,PhysRevB.95.075116,PhysRevB.100.195432}. Regarding the critical exponents $\gamma_{CE}$ and $\nu$, it acquires $0.99$ both at multicritical points and at normal critical points as discussed in the Fig.\ref{mc1}. The values of $\gamma_{CE}$ and $\nu$ are the same throughout the phase diagram which means the scaling law $\gamma_{CE} = \nu$ is obeyed. According to the observation, the model does not exhibit skin effect since there the non-Hermiticity is not direction dependent. This model belongs to the category of non-Hermitian systems complex mass with Lorentz invariance \cite{ge2019topological}\\

\noindent{\textbf{Discussion.}}\label{6}
In this work, we have studied the effect of non-Hermitian factor $\gamma$ in the chemical potential on the topological phases of non-Hermitian extended Kitaev chain with next nearest-neighbor hopping parameters. 
The phase diagram gets modified substantially with the introduction of $\gamma$: specifically, a critical line present in the Hermitian case becomes degenerate isolated critical point in the non-Hermitian model.
We have analyzed the multiple gap-less points in the momentum space interestingly found at the critical and multi-critical points of the non-Hermitian model. We have observed a very distinct and surprising behavior of the real and imaginary parts of the lowest excitation along the line joining the Hermitian and non-Hermitian topological phase transition points. 
We have also analyze the nature of the multi-critical and critical points of the non-Hermitian system. The critical exponents indicates a significant change in the nature of criticality compared to the Hermitian case. In the non-Hermitian case, the value of the dynamical critical exponent $z$ settles down to $0.5$ throughout the parameter space for any value of $\gamma>0$. Also we have observed that the critical exponents $\gamma_{CE}$ and $\nu$ acquires the value 1. Both the values of $\gamma_{CE}$ and $\nu$ remain the same throughout the phase diagram hence obeying the scaling law. 
It will be interesting to see in future if our findings can be established for one dimensional non-Hermitian systems in general. We hope that the current work will lead to many interesting studies to understand non-Hermitian criticality, especially in the topological systems. 

\vspace{0.25cm}
\noindent\textbf{Methods.}\\
\textbf{Zero mode solutions }\label{zms}\\
The model Hamiltonian can be written as,
\begin{equation}
	H_k = \chi_{z}(k) \sigma_z + \chi_{y}(k) \sigma_y,
	\label{1}
\end{equation}
where $ \chi_{z} (k) = 2 \lambda_1 \cos k + 2 \lambda_2 \cos 2k - 2(\mu+ i \gamma),$ and $ \chi_{y} (k) = 2 \lambda_1 \sin k + 2 \lambda_2 \sin 2k.$\\
Substituting the exponential forms of $\cos k$ and $\sin k$, Eq.\ref{1} becomes, 
\begin{dmath}
	H_k = 2 \lambda_1 \frac{1}{2} (e^{-ik} + e^{ik}) + 2 \lambda_2 \frac{1}{2} (e^{-2ik} + e^{2ik} + 2 (\mu+i \gamma))\sigma_z + i 2 \lambda_1 \frac{1}{2} (e^{ik} - e^{-ik}) + 2 \lambda_2 \frac{1}{2}(e^{2ik} - e^{-2ik}) \sigma_y.
	\label{2} 
\end{dmath}  
We replace $e^{-ik} = e^{q}$, Eq.\ref{2} becomes, 
\begin{dmath} 
	H_q =  2 \lambda_1 \frac{1}{2} (e^{q} + e^{-q}) + 2 \lambda_2 \frac{1}{2} (e^{2q} + e^{-2q}) + 2 (\mu+i \gamma) \sigma_z + i  2 \lambda_1 \frac{1}{2} (e^{-q} - e^{q}) + 2 \lambda_2 \frac{1}{2}(e^{-2q} - e^{2q}) \sigma_y. 
	\label{3} 
\end{dmath}   
We make $H_q^2 = 0$ \cite{rahul2021majorana}, to obtain the zero solutions for certain $q$ where $\sigma_z$ and $\sigma_y$ square to 1 or become 0 due to anticommutation. 
\begin{dmath} 
	2 \lambda_1 \frac{1}{2} (e^{q} + e^{-q}) + 2 \lambda_2 \frac{1}{2} (e^{2q} + e^{-2q} + 2 (\mu+i \gamma)) + 2 \lambda_1 \frac{1}{2} (e^{-q} - e^{q}) + 2 \lambda_2 \frac{1}{2}(e^{-2q} - e^{2q}) = 0 
	\label{ezp}
\end{dmath}
Simplifying the Eq.\ref{ezp}, we end up with a quadratic equation,
\begin{dmath} 
	2 \lambda_1 \frac{1}{2} e^{q}  + 2 \lambda_2 \frac{1}{2} e^{2q} + 2 (\mu+i \gamma) +  2 \lambda_1 \frac{1}{2} e^{q} + 2 \lambda_2 \frac{1}{2} e^{2q}  = 0.
	\label{7}
\end{dmath}
Simplifying the Eq.\ref{7} to a quadratic form and substituting $e^q = X$,
\begin{equation} \lambda_2 X^2 +  \lambda_1 X  +  (\mu+i \gamma) = 0.
	\label{8}
\end{equation}
The roots of this quadratic Equation is given by, 
\begin{equation}
	X = \frac{-\lambda_1 \pm \sqrt{\lambda_1^2 + 4 \lambda_2 (\mu+i \gamma)}}{2 \lambda_2} 
	\label{9}
\end{equation}
The roots of Eq.\ref{9} are the solutions of zero modes.
These roots are equated with $1$ and solved for $\lambda_1$ for a given $\lambda_2$, $\mu$ and $\gamma$.
When the solutions are equated to $1$, graphically it means that for given value of $\lambda_2$, the value $\lambda_1$ obtained gives the critical point. In other words, when these solutions are plotted with respect to a unit circle, we are looking at the values of the parameter for which the zero is placed on the unit circle since zeros inside and outside the unit circle corresponds to the topologically non-trivial and trivial phases whereas the zero on the unit circle corresponds to the transition point. Hence by solving the ZMS for different values of $\lambda_2$, one obtains the critical points throughout for all the parameter values. By this method we can obtain the entire phase diagram of the model as shown in Fig.\ref{sch}(II).\\

\begin{figure}[H]
	\centering
	\includegraphics[width=6cm,height=6cm]{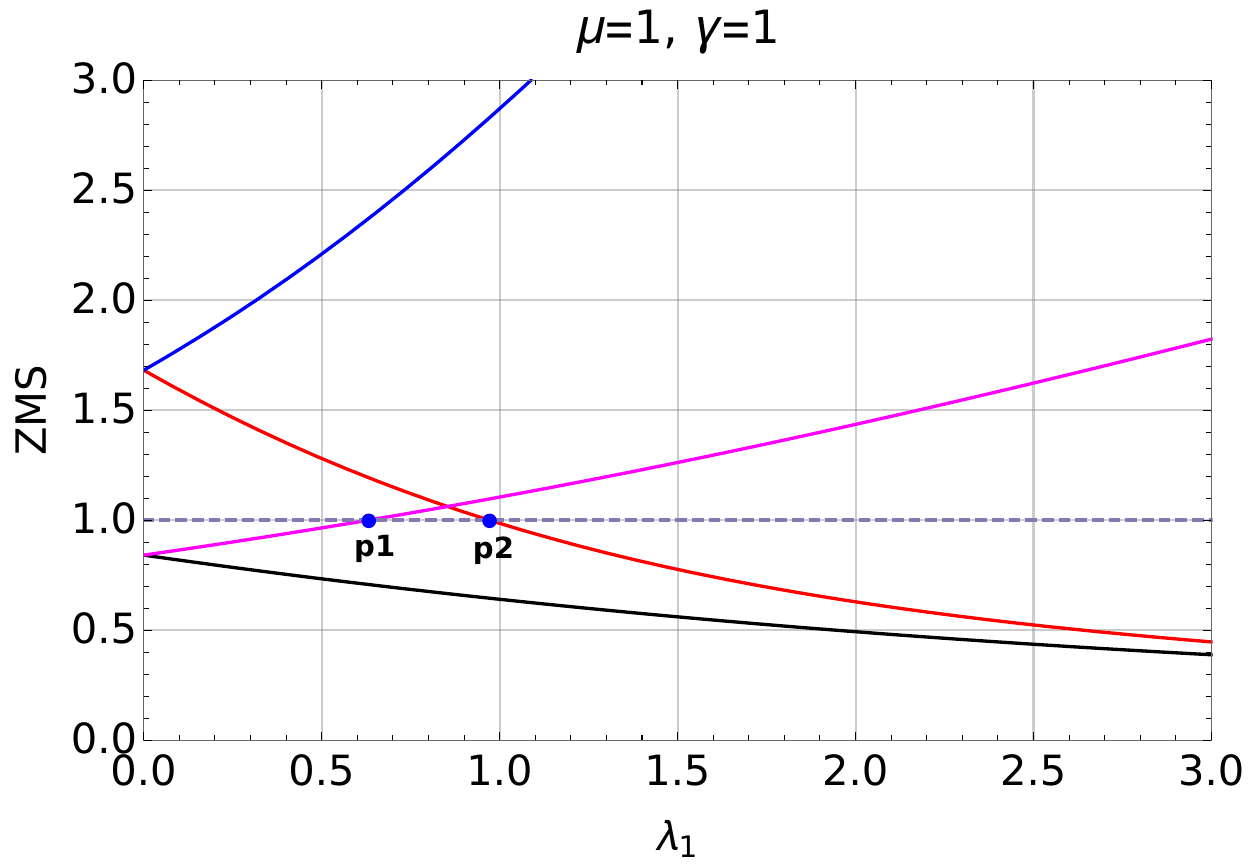}
	\caption{Zero mode solutions plotted with respect to the parameter $\lambda_1$ shows both $W = 0$ to $W = 1$ $(\lambda_2 = 0.5)$ and $W=1$ to $W=2$ $(\lambda_2 = 2.0)$ topological phase transitions.\@ Two blue dots (p1 and p2) represent the transition points.}
	\label{f2}
\end{figure}

Graphically Fig.\ref{f2} corresponds to the process of extracting the critical points in order to plot phase diagram (Fig.\ref{sch}(I)).
In the Fig.\ref{f2}, red and blue curves are the roots representing $W=0$ to $1$ transition. Magenta and black curves are the roots representing $W=1$ to $2$ topological phase transition occurring in $+ \lambda_1$ quadrant of the Fig.\ref{sch}(I). 
To specify the transition point using ZMS, parallel to the x-axis, a reference line (unit line), $y=1$ is drawn which acts in a same way as that of the unit circle drawn to analyze the zeros of a complex function. The point of the intersection between one of the ZMS and the unit line marks the transition point which is represented in blue dots ($p1$ and $p2$) in the Fig.\ref{f2}. \\
Value of the roots greater than 1 corresponds to the non-topological phase whereas less than 1 corresponds to topological phase. Transition points $p1$ and $p2$ marks the critical points of $W=2$ to $1$ and $W=0$ to $1$ topological phase transitions respectively for fixed positive value of $\lambda_2$ ($\lambda_2$ = 0.5, 2.0). Keeping track of these transition points via the ZMS method provides the phase diagram of the model Hamiltonian.\\

\noindent\textbf{Dynamics of gap closing points at criticality}\label{appB}\\
The excitation spectra (see Eq.~\ref{periodb}) of the non-Hermitian model also shows an interesting behavior when compared to the Fig.\ref{peri}(I) and Fig.\ref{peri}(II). 
In Fig.~\ref{peri}(I) and Fig.\ref{peri}(II) the values of $k$ at which the gap closes moves away from $k=0$ point, as we move along the Dashed critical line from positive $\lambda_2$ quadrant to negative $\lambda_2$ quadrant. On contrary as we move along the dotted critical line from negative $\lambda_2$ quadrant to positive $\lambda_2$ quadrant, the values of $k$ where the gap closes moves towards $k=0$ point.
Fig.\ref{MC2} corresponds to the excitation spectra $|E|$ plotted at the multicritical point "d". At this multicritical point, the dynamical critical exponent $z$ acquires the value $0.5$ and the critical exponents $\gamma_{CE}$ and $\nu$ acquires the value $0.99$ which is shown in the Fig.\ref{crit}.\\ 

\begin{figure}
	\centering
	\includegraphics[width=12cm,height=14cm]{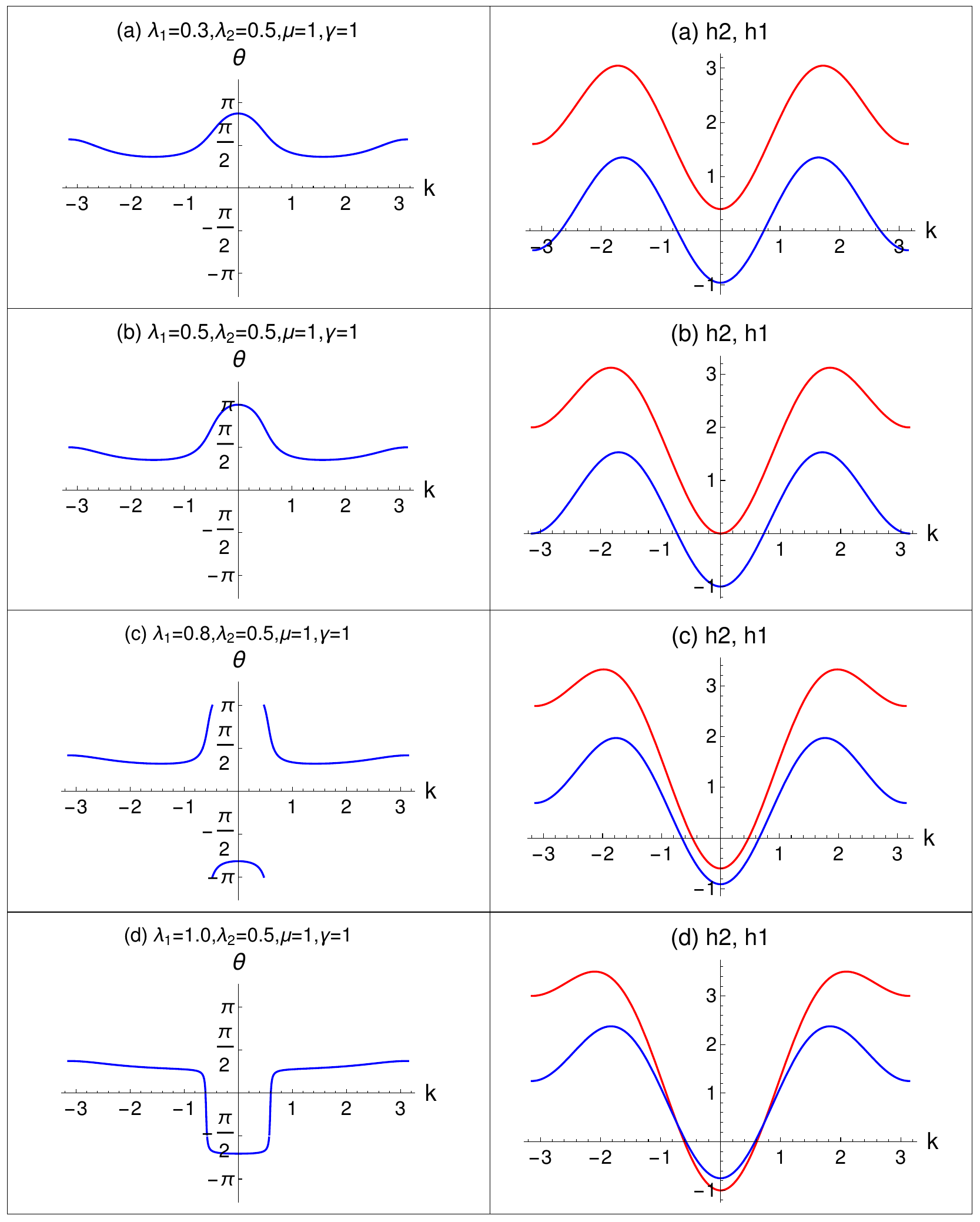}
	\caption{Nature of $\theta$ plotted with respect to $k$ in the left panel. The components of $\theta$ i.e., $h2$ and $h1$ plotted with respect to $k$ in the right panel.}
	\label{theta}
\end{figure}

\noindent\textbf{Mathematical analysis on the behavior of real and imaginary components of the spectrum \label{math}}\\
The dispersion relation is,
\begin{eqnarray}
	E=\pm \sqrt{(\chi_{z} (k))^2 + (\chi_{y} (k))^2},
	\label{di} 
\end{eqnarray} 
where $ \chi_{z} (k) = -2 \lambda_1 \cos k - 2 \lambda_2 \cos 2k + 2(\mu+ i \gamma),$ and $ \chi_{y} (k) = 2 \lambda_1 \sin k + 2 \lambda_2 \sin 2k$.
Expanding the dispersion relation (Eq.~\ref{di}) and writing it in terms of $x+i y$ form, we get,
\begin{eqnarray}
	E=\pm r(\cos\frac{\theta}{2}+ i \sin\frac{\theta}{2}),
	\label{en}
\end{eqnarray}
where \begin{dmath}
	r = 2 (4 \gamma^2 (-\mu +\lambda_1 \cos k + \lambda_2 \cos 2k)^2 + (-\gamma^2 + \lambda_1^2 + \lambda_2^2 + \mu^2 + 2 \lambda_1 (\lambda_2 - \mu) \cos k - 2 \lambda_2 \mu \cos 2k)^2)^{\frac{1}{4}}
\end{dmath} and
\begin{equation}
	\theta = \arctan\left( \frac{h1}{h2}\right),
	\label{th}
\end{equation}
where $h1 = -\gamma^2 + \lambda_1^2 + \lambda_2^2 + \mu^2 + 2 \lambda_1 (\lambda_2 - \mu) \cos k - 2 \lambda_2 \mu \cos 2k$ ,
$h2= 2 \gamma \mu - 2 \gamma \lambda_1 \cos k - 2 \gamma \lambda_2 \cos 2k$.\\
\noindent The behavior of $\theta$ is studied with respect to $k$ to understand the real imaginary components of the eigenvalues which has a $\theta$ factor in them when written in the complex form.\\
From the Fig.~\ref{theta}, $\theta$ is a $\gamma$ induced argument which dictates the behavior of real and imaginary components. Fig.~\ref{theta} comprises of left and right panel. In the left panel, $\theta$ is studied with respect to $k$ and in right panel, $h1$ and $h2$ of $\theta$ (Eq.~\ref{th}) are studied with respect to $k$.
In the right panel, in plot~(a), $h2$ is positive and $h1$ is negative and in plot~(b) $h2$ is zero and $h1$ is negative at $k=0$. Hence $\theta$ at $k=0$ takes the value $\theta + \pi$ which is reflected in the plot~(a) and (b) of left panel. In plot~(c) and (d) of right panel, both $h2$ and $h1$ components are negative at $k=0$ and hence $\theta$ takes the value $\theta - \pi$ which is also reflected in the plot~(c) and (d) of left panel.\\
When either one of $h2$ or $h1$ of $\theta$ are negative and other remain positive, the argument takes the value $\theta + \pi$. Since $\theta$ is negative and less than $\pi$, it is positive.\\
When both $h2$ and $h1$ are negative at $k=0$, argument takes the value $\theta - \pi$ and since $\theta < \pi$, it becomes negative as shown in the plot~(c) and (d), right panel of Fig.\ref{theta}. This inherent nature of $\theta$ is responsible for the behavior of real and imaginary components of eigenvalues that we observe in previous sections.
Investigating the Eq.~\ref{th} in detail with the help of $\theta$, we come across cosine and sine terms for real and imaginary components respectively. When $\theta$ is negative, real component consist of cosine term which is an even function. Hence, the real component of eigenvalue never crosses from positive to negative quadrant like the imaginary component does. Imaginary component consist of sine term which is an odd function and hence the imaginary component makes a jump from negative to positive values and vice versa.

\noindent\textbf{Analysis on the imaginary excitation spectrum \label{levelc}}\\
Excitation spectrum of the model Hamiltonian (Eq.~\ref{APS}) given as, 
\begin{dmath}
	E_{\mu+i \gamma}=\pm \sqrt{(\chi_{z} (k))^2 + (\chi_{y} (k))^2},
	\label{es1}
\end{dmath}
where $ \chi_{z} (k) = -2 \lambda_1 \cos k - 2 \lambda_2 \cos 2k + 2(\mu+ i \gamma),$ and $ \chi_{y} (k) = 2 \lambda_1 \sin k + 2 \lambda_2 \sin 2k.$ 
Expanding the complex term into real and imaginary components, Eq.~\ref{es1} is rewritten as, 
\begin{dmath} 
	E_{\mu+i \gamma}=\pm r (\cos\frac{\theta}{2}+ i \sin\frac{\theta}{2}),
	\label{g1}
\end{dmath}
where \begin{dmath}
	r = 2 (4 \gamma^2 (-\mu +\lambda_1 \cos k + \lambda_2 \cos 2k)^2 + (-\gamma^2 + \lambda_1^2 + \lambda_2^2 + \mu^2 + 2 \lambda_1 (\lambda_2 - \mu) \cos k - 2 \lambda_2 \mu \cos 2k)^2)^{\frac{1}{4}}
\end{dmath} and
\begin{dmath} 
	\theta = \arg[(i \gamma + \mu - \lambda_1 \cos k - \lambda_2 \cos 2k)^2 + (\lambda_1 + 2 \lambda_2 \cos k )^2 \sin k^2].\label{g2}
\end{dmath} 
Considering only the imaginary part of the spectrum from Eq.~\ref{g1}, 
\begin{dmath}  
	Im:E_{\mu+i\gamma} = \pm r (\sin[\frac{1}{2}\arg[(i \gamma + \mu - \lambda_1 \cos k - \lambda_2 \cos 2k)^2  + (\lambda_1 + 2 \lambda_2 \cos k )^2 \sin k^2]]) 
	\label{d1}
\end{dmath}
From a careful analysis, one can see that positive and negative imaginary roots corresponds to the imaginary roots of the excitation spectrum when the non-Hermitian term is replaced with $-i \gamma$. The excitation spectra with $-i\gamma$ is,
\begin{dmath}
	E_{\mu-i\gamma}=\pm \sqrt{(\chi_{z} (k))^2 + (\chi_{y} (k))^2},
	\label{es2}
\end{dmath}
where $ \chi_{z} (k) = -2 \lambda_1 \cos k - 2 \lambda_2 \cos 2k + 2(\mu- i \gamma),$ and $ \chi_{y} (k) = 2 \lambda_1 \sin k + 2 \lambda_2 \sin 2k.$\\  
Expanding the complex part of the dispersion relation and focusing only on the imaginary part, we get, 
\begin{dmath} 
	Im:E_{\mu-i\gamma} = \pm r ( \sin[\frac{1}{2}\arg[(i \gamma - \mu + \lambda_1 \cos k + \lambda_2 \cos 2k)^2 + (\lambda_1 + 2 \lambda_2 \cos k )^2 \sin k^2]]).
	\label{d2} 
\end{dmath}
\begin{figure}[H]
	\centering
	\includegraphics[scale=0.4]{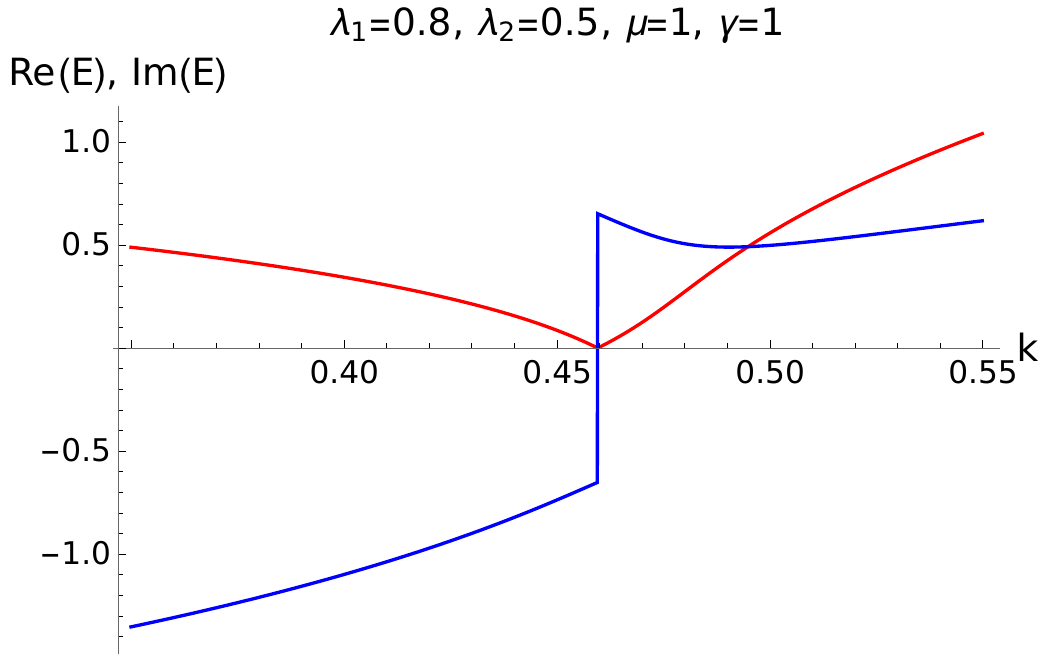}
	\caption{Real (in red) and imaginary (in blue) components of the excitation plotted at a critical point.}
	\label{r2}
\end{figure}
The observation of imaginary component changing sign in the Fig.\ref{peri}(III) is also observed in the excitation spectra Fig.\ref{r2} under periodic boundary condition. During the sign change of imaginary component, the real component becomes zero which is observed from both Fig.\ref{peri}(III) and Fig.\ref{r2}. Although it is not physically consistent to draw a comparison between the open boundary setting and periodic boundary setting, the change of sign of imaginary component when the real component becomes zero is commonality is observed in the behavior of the real imaginary components. This behavior is clearly studied using the Eqs. \ref{d1} and .\ref{d2}. It is shown here that, the change in sign of the imaginary component corresponds to the change in the sign of the non-Hermitian factor $\gamma$ $(\pm i\gamma)$. The physical aspect of this behavior is $+i\gamma$ and $-i\gamma$ corresponds to the gain and loss of the system with the on-site local bath that is attached. 

\section*{Acknowledgments}
The authors would like to acknowledge DST (CRG/2021/000996) for the funding and RRI library for the books and journals. NR acknowledges financial support from the Indian Institute of Science through the IOE-PDF fellowship scheme. Authors would like to acknowledge Dr. Chung Hua Lee for useful interaction.
\section*{Author contributions}
S.S. identified the problem, S.R and N.R solved the problem and wrote the manuscript under the guidance of S.S. Y.R.K and R.K.R helped in analyzing the results. 
All the authors reviewed the manuscript.\\
\section*{Competing interests}
The authors declare no competing interests.
\section*{Additional information}
Correspondence and requests for materials should be addressed to S.S.
\section*{Data availability}
All data generated or analyzed during this study are included in this published article.

\end{document}